\documentclass[prl,aps,twocolumn,showpacs,superscriptaddress]{revtex4}

\usepackage{dcolumn}
\usepackage{bm}
\usepackage{amsmath,amssymb}
\usepackage{graphicx}
\usepackage{hyperref}

\newcommand{\ket}[1]{\left|#1\right>}
\newcommand{\bra}[1]{\left<#1\right|}
\newcommand{\braket}[2]{\left<#1|#2\right>}
\newcommand{\IDM}[0]{\text{IDM}}

\begin{document}

\title{Decoherence in Attosecond Photoionization} 

\author{Stefan Pabst}
\affiliation{Center for Free-Electron Laser Science, DESY, Notkestrasse 85, 22607 Hamburg, Germany}
\affiliation{Department of Physics,University of Hamburg, Jungiusstrasse 9, 20355 Hamburg, Germany}

\author{Loren Greenman}
\affiliation{Department of Chemistry and The James Franck Institute, The University of Chicago, Chicago, Illinois 60637, USA}

\author{Phay J. Ho}
\affiliation{Argonne National Laboratory, Argonne, Illinois 60439, USA}  

\author{David A. Mazziotti}
\affiliation{Department of Chemistry and The James Franck Institute, The University of Chicago, Chicago, Illinois 60637, USA}
	
\author{Robin Santra}
\thanks{Corresponding author}
\email{robin.santra@cfel.de}
\affiliation{Center for Free-Electron Laser Science, DESY, Notkestrasse 85, 22607 Hamburg, Germany}
\affiliation{Department of Physics,University of Hamburg, Jungiusstrasse 9, 20355 Hamburg, Germany}

\date{\today}


\begin{abstract}
The creation of superpositions of hole states via single-photon ionization using attosecond extreme-ultraviolet pulses is studied with the {\it time-dependent configuration interaction singles} (TDCIS) method.
Specifically, the degree of coherence between hole states in atomic xenon is investigated.
We find that interchannel coupling not only affects the hole populations, it also enhances the entanglement between the photoelectron and the remaining ion, thereby reducing the coherence within the ion. 
As a consequence, even if the spectral bandwidth of the ionizing pulse exceeds the energy splittings among the hole states involved, perfectly coherent hole wave packets cannot be formed. 
For sufficiently large spectral bandwidth, the coherence can only be increased by increasing the mean photon energy. 
\end{abstract} 

\maketitle


The typical time scale of electronic motion in atoms, molecules, and condensed matter systems 
ranges from a few attoseconds ($1~\text{as} = 10^{-18}$~s) to tens of femtoseconds ($1~\text{fs} = 10^{-15}$~s) 
\cite{KrIv-RMP81,Zewail_JCPA104,CaMu-Nature449-2007}. 
In the last decade the remarkable progress in high harmonic generation 
\cite{DoWh-PRL102-2009,DuSm-NaturePhys2-2006,LoVa-PRL94-2005,GiPa-PRL92-2004,ScYa-PRL70-1993} made it possible to \
generate attosecond pulses as short as 80~as \cite{GoSc-Science320}.  Attosecond pulses have opened the door to 
real-time observations of the most fundamental processes on the atomic scale \cite{KrIv-RMP81,PfAb-CPL463-2008}.
For instance, the generation of attosecond pulses was utilized to determine spatial structures of molecular orbitals 
\cite{haessler_attosecond_2010}; an interferometric technique using attosecond pulses was used to characterize
attosecond electron wave packets \cite{MaRe-PRL105-2010}; and attosecond pulse trains \cite{SiHe-PRL104-2010}
and isolated attosecond pulses \cite{SaKe-Nature465-2010}, in combination with an intense few-cycle infrared pulse, 
enabled the control of electron localization in molecules.  Attosecond technology demonstrated the ability to follow, 
on a subfemtosecond time scale, processes such as photoionization \cite{SkTz-PRL105-2010}, Auger decay \cite{drescher_time-resolved_2002}, 
and valence electron motion driven by relativistic spin-orbit coupling \cite{goulielmakis_real-time_2010}. 
Furthermore, the availability of attosecond pulses fuelled a broad interest in exploring charge transfer dynamics following 
photoexcitation or photoionization \cite{SaKe-Nature465-2010}. 

In this Letter, we analyze the creation of hole states via single-photon ionization using a single extreme-ultraviolet attosecond pulse. 
We investigate the impact of the freed photoelectron on the remaining ion and demonstrate that the interaction between the photoelectron 
and the ion cannot be neglected for currently available state-of-the-art attosecond pulses.
In particular, the interchannel coupling of the initially coherently excited hole states greatly enhances the entanglement between the photoelectron and
the ionic states. Interchannel coupling is mediated by the photoelectron and mixes
different ionization channels, i.e., hole configurations, with each other. Consequently, the degree
of coherence among the ionic states is strongly reduced, making it impossible to describe the subsequent charge
transfer in the ion with a pure quantum mechanical state. Experiments on photosynthetic systems
\cite{LeCh-Science316-2007,SaMo-NatPhys6-2010,CoWo-Nature463-2010,HarFi-PNAS107-2010} have revealed
a correlation between highly efficient energy transport and coherent dynamics in molecules (nuclear and electronic dynamics in this case). 
Similarly, high degrees of coherence in nonstationary hole states
may be necessary for efficient charge transport within molecules.
   
In the last decade, much work has been done in the realm of hole migration 
\cite{breidbach_migration_2003,NeRe_NJoP10-2008,kuleff_ultrafast_2010}. 
It was shown that electronic motion can be triggered solely by electron correlation \cite{breidbach_migration_2003}. 
Charge transfers mediated by electronic correlations typically take place in a few femtoseconds and are thus faster than electronic dynamics 
initiated by nuclear motion \cite{KrKl-ApplPhysA88-2007,MuRe_PhysScripta80-2009}.
Recent experiments \cite{WeSc_JPCA42,schlag_distal_2007} have demonstrated that electronically excited ionic states can modify 
site-selective reactivity within tens of femtoseconds, making hole migration processes a promising tool to control chemical reactions.
Up to now, theoretical calculations \cite{breidbach_migration_2003,kuleff_ultrafast_2010} investigating hole migration phenomena 
have neglected the interaction between the parent ion and the photoelectron and assumed a perfectly coherent hole wave packet.
As long as the photoelectron departs sufficiently rapidly from the parent ion, this assumption is appropriate \cite{CdDo_AdvChemPhys65}. 
However, for attosecond pulses with large spectral bandwidths, the enhanced production of slow
photoelectrons will affect (mainly via interchannel coupling) both the final hole populations and
the coherence among these hole states.  Furthermore, recent results in high harmonic spectroscopy suggest that interchannel coupling 
may be the missing link to understand hole dynamics occurring in high harmonic generation processes before the ejected electron recombines 
with the parent ion \cite{MaHi-PRL104-2010}.

We investigate the creation of hole states via attosecond photoionization using the implementation of the time-dependent configuration-interaction singles 
(TDCIS) approach described in Ref. \cite{GrHo_PRA82} (see also \cite{ScSm-JCP126-2007,KlSa-JCP131-2009}). 
TDCIS allows us to study ionization dynamics beyond the single-channel 
approximation and to understand systematically the relevance of interchannel coupling in the hole
creation process. The TDCIS wave function for the entire system is
\begin{eqnarray}
  \label{eq:1}
  \ket{\Psi(t)}
  =
  \alpha_0(t)\,\ket{\Phi_0}
  +
  \sum_{a,i} \alpha^a_i(t)\,\ket{\Phi^a_i},
\end{eqnarray}
where $\ket{\Phi_0}$ is the Hartree-Fock ground state and $\ket{\Phi^a_i}=\hat c^\dagger_a \hat c_i \ket{\Phi_0}$ is a one-particle--one-hole 
excitation ($\hat c^\dagger_a$ and $\hat c_i$ are creation and annihilation operators for an electron in orbitals $a$ and $i$, respectively).
The corresponding coefficients $\alpha_0(t)$ and $\alpha^a_i(t)$, respectively, are functions of time and describe the dynamics of the system. 
Throughout, indices $i,j,$ are used for occupied orbitals in $\ket{\Phi_0}$; indices $a,b,$ stand for unoccupied orbitals.
We focus our discussion on the case where single-photon ionization is the dominant effect and higher
order processes can be neglected.  Our model system is atomic xenon. The corresponding Hamiltonian (neglecting spin-orbit coupling) is
\begin{subequations}
\label{eq:2}
\begin{eqnarray}
  \label{eq:2.1}
	\hat H(t)
	&=&
	\hat H_0 + \hat H_1 + E(t)\,\hat z ,
\end{eqnarray}
where $E(t)$ is the electric field, $\hat z$ the dipole operator, and $\hat H_0$ is the mean-field Fock operator, which is diagonal 
with respect to the basis used in Eq.~(\ref{eq:1}). The residual Coulomb interaction,
\begin{eqnarray}
  \label{eq:2.2}
	\hat H_1
	&=&
	\hat V_c - \hat V_\text{MF},
\end{eqnarray}
\end{subequations}
is defined such that $\hat H_0 + \hat H_1$ gives the exact nonrelativistic Hamiltonian for the electronic system in the absence of
external fields ($\hat V_c$ is the electron--electron interaction).
We study the impact of different approximations for $\hat H_1$ on the hole state as follows. 
The {\it Coulomb-free} model, the simplest approximation, removes the residual Coulomb interaction ($\hat H_1=0$) between the excited electron
and the parent ion. In this approximation, the excited electron always sees a neutral atom via the $\hat V_\text{MF}$ potential \cite{SzaboOstlund-book-1996}. 
A more realistic approximation is the {\it intrachannel} model including direct and exchange contributions of the Coulomb interaction 
only within a given channel. In this second model, the excited electron can only interact with the
occupied orbital from which it originates. Interactions between different occupied orbitals are neglected, i.e. we set $\bra{\Phi^a_i}\hat H_1\ket{\Phi^b_j}=0$~ for $i\neq j$.
The third and final model describes the Coulomb interaction exactly within the TDCIS framework.  We refer to this as the {\it full} model. 
Note that the exact nonrelativistic Hamiltonian $\hat H_0 + \hat H_1$ is diagonal with respect to the ionic one-hole states
$\ket{\Phi_i}=\hat c_i \ket{\Phi_0}$.  In the full model, the photoelectron can couple the hole states, as $\hat H_1$ in the particle-hole
space is not diagonal with respect to the hole index (i.e., $\bra{\Phi^a_i}\hat H_1\ket{\Phi^b_j}$ generally differs from zero).
This type of photoelectron-mediated interaction is called interchannel coupling \cite{Starace:80}.  As a consequence, in the full model
the hole index is not a good quantum number, whereas in the Coulomb-free and intrachannel models, excited eigenstates of $\hat H_0 + \hat H_1$
are characterized by a well-defined hole index.
To describe the hole states of the remaining ion, we employ the ion density matrix \cite{GrHo_PRA82}
\begin{eqnarray}
  \label{eq:3}
  \hat\rho^\IDM_{i,j}(t)
  &=&
  \text{Tr}_a[\ket{\Psi(t)}\bra{\Psi(t)}]_{i,j}
	=
	\sum_a \braket{\Phi^a_i}{\Psi(t)}\braket{\Psi(t)}{\Phi^a_j},
\end{eqnarray}
where $\text{Tr}_a$ stands for the trace over the photoelectron. 
The properties of the ion density matrix can be measured using attosecond transient absorption spectroscopy \cite{goulielmakis_real-time_2010}.
A description of the cationic eigenstates in terms of one-hole configurations is a physically meaningful approximation for noble-gas atoms such as xenon \cite{BuSa-JCP119-2003}.

\begin{figure}[!ht]
\begin{center}
  \includegraphics[clip,width=\linewidth]{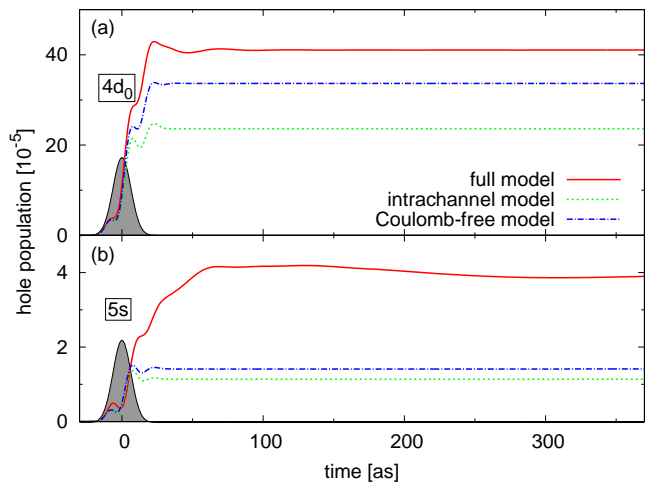}
  \caption{(color online) The $4d_0$ [panel (a)] and $5s$ [panel (b)] hole populations of xenon as a function of time are shown for 
three different residual Coulomb interaction approximations: (1) the full model (red solid line), (2) the intrachannel 
model (green dotted line), and (3) the Coulomb-free model (blue dash-dotted line). 
The attosecond pulse has a peak field strength of 25~GV/m, a pulse duration of 20~as, a (mean) photon energy of 136~eV, and is centered at $t=0$~as.}
  \label{fig.1}
\end{center}
\end{figure}
In Fig. \ref{fig.1} the hole populations $\rho^\IDM_{5s,5s}(t)$ and $\rho^\IDM_{4d_0,4d_0}(t)$ of the xenon $5s$ and $4d_0$ orbitals, 
respectively, are shown for all three interaction models ($4d_0$ stands for the $4d$ orbital with $m=0$).
The ionizing, gaussian-shaped attosecond pulse is linearly polarized and has a peak field strength
of 25~GV/m, a pulse duration of $\tau=20$~as, and a (mean) photon energy of $\omega_0=136$~eV.
The hole dynamics of the Coulomb-free and intrachannel models are alike. In both cases, the
population is constant after the pulse, since the hole index is a good quantum number within these
models. The extension to the exact Coulomb interaction changes the
situation. Interchannel coupling causes the hole populations to remain nonstationary as long as the
photoelectron remains close to the ion. As the distance between the photoelectron and the ion
increases, the interchannel coupling weakens and the populations $\rho^\IDM_{i,i}(t)$ become
stationary (see Fig.~\ref{fig.1}). We confine our discussion to the first hundreds of attoseconds
after the pulse, allowing us to neglect decay processes, which  start to take place after a few femtoseconds.

As we will see in the following, interchannel coupling not only affects the hole populations but also the coherence between the created hole 
states.  The degree of coherence between $\ket{\Phi_i}$ and $\ket{\Phi_j}$ is given by
\begin{eqnarray}
  \label{eq:4}
  g_{i,j}(t)=\frac{|\rho^\IDM_{i,j}(t)|}{\sqrt{\rho^\IDM_{i,i}(t)\rho^\IDM_{j,j}(t)}}.
\end{eqnarray}  
Totally incoherent statistical mixtures result in $g_{i,j}(t)=0$.
The fact that the density matrix is positive semidefinite implies the Cauchy-Schwarz
relations $|\rho^\IDM_{i,j}(t)|^2 \leq \rho^\IDM_{i,i}(t)\rho^\IDM_{j,j}(t)$, which bound the
maximum achievable (perfect) coherence ($g_{i,j}(t)=1$). To investigate the effect of
interchannel coupling on the coherence between the orbitals $4d_0$ and $5s$ in xenon, we restrict
the definition of the $4d_0$ hole population to the events where the photoelectron has angular
momentum $l=1$. The other possible angular momentum for the $4d_0$ photoelectron, $l=3$, does not
contribute to the coherence, since the photoelectron from $5s$ can only have $l=1$.  For a similar
reason, it is impossible to create a coherent superposition of $5p$ and $5s$ (or $4d$) hole states
via one-photon absorption in the electric dipole approximation.

\begin{figure}[!ht]
\begin{center}
  \includegraphics[clip,width=\linewidth]{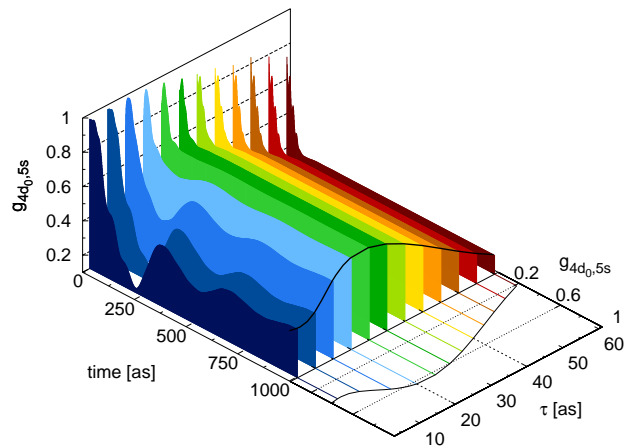}
  \caption{(color online) The time evolution of the coherence between the $4d_0$ and $5s$ 
  hole states in xenon is shown for the full Coulomb interaction model. The photon energy is 136~eV
  and the pulse duration varies from 5--60~as.}
  \label{fig.2}
\end{center}
\end{figure}

Figure \ref{fig.2} illustrates the time evolution of the coherence between $4d_0$ and $5s$ in xenon for different pulse durations 
and fixed photon energy ($\omega_0=136$~eV). Here, we use the full interaction model.
Directly after the ionizing pulse is over, the initial degree of coherence (at $t\approx 0$~as)
rises with decreasing pulse duration, i.e., increasing spectral bandwidth, and converges to a value close to
unity.
(The difference of the ionization potentials, $\varepsilon_{5s}-\varepsilon_{4d_0}$, is $\approx
50$~eV.)
At $t\approx 0$~as, the photoelectron is still in immediate contact with the parent ion.
Therefore, the coherence properties of the system of interest---the parent ion---are affected by its
interaction with the bath represented by the photoelectron.
The system--bath interaction leads to a reduction in the coherence of the system
\cite{Breuer-Bk-OpenQuantSyst-2002}, which can be seen by the rapid drops in all curves in Fig.
\ref{fig.2} within tens of attoseconds after the pulse.
With time, as the photoelectron departs from the ion, the Coulomb
(``system-bath'') interaction becomes less important and the coherence converges to a stationary value.
The maximum for this stationary value is obtained with a 25~as pulse ($g_{4d_0,5s}\approx0.6$).
For pulses shorter than 25~as, oscillations in $g_{4d_0,5s}$ occur that persist for hundreds of
attoseconds, and the final degree of coherence reached falls below 0.6.
The spin-orbit dynamics associated with the fine-structure within the $4d$ shell is slow in comparison
to the time scale of the decoherence between $4d_0$ and $5s$, and is, therefore, not considered here.

\begin{figure}[!ht]
\begin{center}
  \includegraphics[clip,width=\linewidth]{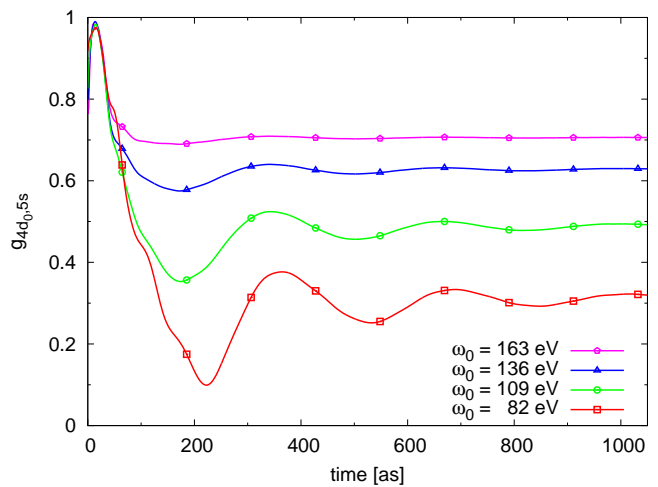}
  \caption{(color online) The time evolution of the coherence between the $4d_0$ and $5s$ hole states, calculated with the full Coulomb 
interaction model, is shown for different photon energies. The pulse duration is in all cases 20~as.}
  \label{fig.3}
\end{center}
\end{figure}

We see in Fig.~\ref{fig.3} that when holding the pulse duration fixed ($\tau=20$~as), 
the degree of coherence rises with increasing $\omega_0$.  The magnitude of the oscillations
decreases as the final coherence (at $t\approx 1$~fs) increases. This trend indicates less
system-bath interactions occur with higher photoelectron energies keeping the degree of coherence among the hole states high. 

\begin{figure}[!ht]
\begin{center}
  \includegraphics[clip,width=\linewidth]{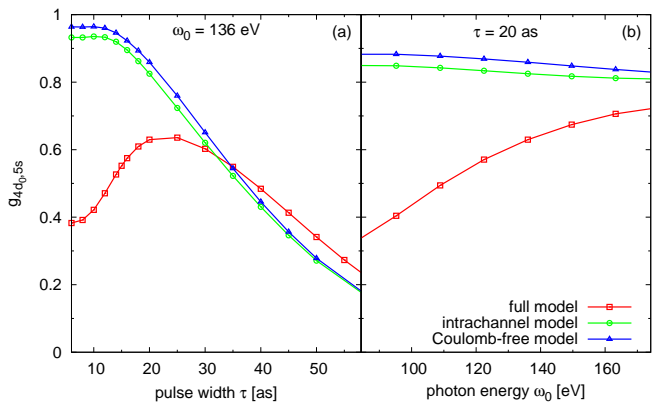}
  \caption{(color online) The dependence of the coherence between the $4d_0$ and $5s$ hole states as
  function of the pulse duration (a) and as function of the photon energy (b) are shown for
  all three interaction approximations.}
  \label{fig.4}
\end{center}
\end{figure}

In Fig.~\ref{fig.4} we compare the impact of the different Coulomb approximations on the final
coherence. The drops in coherence that occur for the full model for short pulses
[Fig.~\ref{fig.4}(a)] and low photon energies [Fig.~\ref{fig.4}(b)] cannot be seen in the Coulomb-free 
and intrachannel models---which both neglect interchannel coupling. Hence, the decay of coherence is
solely driven by the interchannel coupling due to the slow photoelectron. As a comparison to the 
Coulomb-free model shows, intrachannel coupling affects the coherence in an insignificant way.
In the limit of long pulse durations (small spectral bandwidths), the coherence vanishes for all models, since photoelectrons 
from the $4d_0$ and $5s$ become energetically distinguishable and cannot contribute to a coherent statistical mixture of hole states.
The slight drop in the coherence for the Coulomb-free and intrachannel models with increasing $\omega_0$ [Fig.~\ref{fig.4}(b)]
is related to the reduced factorizability of the numerator of Eq.~(\ref{eq:4}).  In contrast, the trend in the full model for 
increasing $\omega_0$ is dominated by the gain in coherence due to higher photoelectron energy resulting in less system-bath interaction.

In conclusion, we demonstrated that the coherence of the ionic states produced via attosecond photoionization is not solely determined
by the bandwidth of the ionizing pulse, but greatly depends on the kinetic energy of the photoelectron, which can be controlled by the 
(mean) photon energy.  Interchannel coupling leads to an enhanced entanglement between the photoelectron and the parent ion resulting 
in a reduced coherence in the ionic states.  This reduction can be mitigated with higher photon energies, thereby sacrificing high 
photon cross sections and the possibility of controlling independently the relative populations of
the various hole states in the statistical mixture. 

Our results have far-reaching consequences beyond the atomic case.
Molecules will be even more strongly affected by interchannel coupling due to the reduced symmetry and smaller energy splittings
between the cation many-electron eigenstates. Interchannel coupling is also likely to be significant
for inner-valence hole configurations in molecules, which show strong mixing to configurations outside the TDCIS model space. 
The present study suggests that interchannel coupling accompanying
the hole creation process will affect attosecond experiments investigating charge transfer processes
in photoionized systems. The control of decoherence requires widely tunable attosecond
sources, thus offering a new opportunity for x-ray free-electron lasers
\cite{ZhFa-PRL92-2004}.

\acknowledgments
P.J.H. was supported by the Office of Basic Energy Sciences, U.S. Department of Energy
under Contract No. DE-AC02-06CH11357. L.G. thanks Martha Ann and Joseph A. Chenicek and
their family. D.A.M. gratefully acknowledges the NSF, the Henry-Camille
Dreyfus Foundation, the David-Lucile Packard Foundation, and the Microsoft Corporation for their
support.



\end{document}